\documentclass[12pt,preprint]{aastex}

\def\ls{{_<\atop^{\sim}}}
\def\gs{{_>\atop^{\sim}}}

\slugcomment{}
\shorttitle{Smoking Quasars: a New Source for Cosmic Dust}
\shortauthors{M. Elvis et al.}

\begin{document}


\title{Smoking Quasars: a New Source for Cosmic Dust}

\author{Martin Elvis, Massimo Marengo and  Margarita Karovska}
\affil{Harvard-Smithsonian Center for Astrophysics, 60 Garden St.,
Cambridge, MA 02138}
\email{elvis,mmarengo,karovska@cfa.harvard.edu}


\begin{abstract}
Although dust is widely found in astrophysics, {\em forming} dust
is surprisingly difficult.  The proper combination of low
temperature ($<$2000~K) and high density is mainly found in the
winds of late-type giant and supergiant stars which, as a result,
are the most efficient sources of dust known.  Dust ejected from
these stars into the interstellar medium has multiple important
effects, including obscuring background objects and enhancing
star formation.
We show here that quasars are also naturally copious producers of
dust, if the gas clouds producing their characteristic broad
lines are part of an outflowing wind. This offers an explanation
for the strong link between quasars and dust, for the heavy
nuclear obscuration around many quasars and introduces a new
means of forming dust at early cosmological times.
\end{abstract}



\section{Introduction}\label{sec-intro}

It is widely believed that the quasar environment is hostile to
dust, so it is striking that quasars contain large amounts of
dust, even at high redshifts (z$>$4, Omont et al. 2001, Priddey
\& McMahon 2001) when the Universe is only 1-2~Gyr old
(H$_0$=65~km~s$^{-1}$~Mpc$^{-1}$).  How dust can form so early is
a significant problem (Edmunds \& Eales 1998). Normally the dust
in quasars is assumed to arise from independent processes
unconnected with the quasar.  However an intrinsic origin is
hinted at: the outer radius of the region producing the strong
broad emission lines (BELs) is typically coincident with the dust
sublimation radius (Clavel, Wamsteker \& Glass 1989). We show
that the free expansion of the clouds producing the quasar BELs
produces conditions conducive to the formation of dust, making
quasars one of the few sources of cosmic dust.

Carbon rich dust does form in similarly unfavorable-seeming
environments, such as the UV-bright Wolf-Rayet and R Corona
Borealis stars. These hot giants have effective photospheric
temperatures and wind densities which are strikingly similar to
the initial conditions of the BEL clouds ($T_{eff} \ge 2 \cdot
10^4$~K, Sedlmayr 1997).

The gas producing the BELs lies in a large number of small
`clouds' at temperatures of 10$^4$~K and densities of
10$^9$-10$^{11}$~cm$^{-3}$ (Osterbrock 1989).  BELs are Doppler
broadened to a few percent of the speed of light ($\sim
3000-15,000$~km~s$^{-1}$). As yet though, the kinematics of the
BEL clouds - whether they are infalling, in bound orbits, or in
an outflowing wind - is not well established (Peterson
1997). Moreover, the issue of how to prevent these clouds from
dispersing has been problematic. Pressure confinement by a hotter
surrounding medium would seem straightforward, but appeared to
suffer unsurmountable problems, in a simple spherical geometry
(Mathews 1986).

Outflowing winds of highly ionized material are common in quasars
(Arav, Shlosman \& Weymann 1997), so a similar flow pattern for
BEL clouds is a reasonable possibility.  Elvis (2000) showed how
such a wind, if highly non-spherical, can solve the confinement
problems of the BEL clouds by assuming that they are a cool
phase in equilibrium with a warmer (10$^6$~K) wind medium.  With
this model a large number of other puzzling features of quasar
phenomenology seem to fall into place.  The fate, however, of the
outflowing BEL clouds constantly being ejected from the quasar
were not considered by Elvis. We examine that fate
here in a manner that in fact applies to any model in which the
BEL clouds move outward. We find that dust creation is a natural
consequence.

In any outflow model that begins with the BEL region initially in
pressure equilibrium with a surrounding warmer medium, the
divergence of the outflowing warm wind (even if only at the sound
speed of the warm confining medium, $\sim$100~km~s$^{-1}$) will
rapidly take the system out of pressure balance
\footnote{Evidence from reverberation mapping currently suggests
that the BELR is not in overall expansion (Peterson 1997). This
is expected in the Elvis (2000) picture, where the BELR is
initially rotating and rising in a cylinder, and is only later
blown outward, e.g. by radiation pressure, to form a radial wind;
the BEL gas is shielded in the radial wind region from much of
the quasar ionizing continuum.}
.The BEL clouds will then begin to expand, limited by their sound
speed (initially $\sim$10~km~s$^{-1}$), and will cool to
temperatures at which dust will form, if the pressure is still
sufficiently high and if the dust forming species are
available. This process resembles that which makes smoke in
terrestrial settings. Quasars are then dusty because they
themselves create dust, which may resemble soot: they are
`smoking quasars'.

A full treatment of dust formation generally requires coupling
the dust forming medium hydrodynamics with the full set of dust
condensation chemistry equations (Sedlmayr 1997). Such an
exercise is limited by our knowledge of the highly non-linear
dust condensation chemical paths, and by the uncertainties
related to the role of non-equilibrium chemistry. We therefore
use a simple comparison between AGN and cool star atmospheres,
to derive reasonable estimates for the conditions of dust
formation in the BEL clouds.

\section{The dust formation window}\label{window}

The general scenario for dust formation is based on the concept
of the ``dust formation window''. Effective dust condensation
seems to take place whenever a chemically enriched medium has a
sufficiently low temperature, and a large enough density, to
allow dust grain condensation. The amount of dust produced, the
chemical composition and the final size of the grains, depend on
the length of time over which the conditions remain favorable.

The typical dust grains encountered in the interstellar medium
have condensation equilibrium temperatures in the range
600--1400~K or somewhat higher (Frenklach \& Feigelson
1989). These conditions are generally met in the cool atmospheres
of late type giants and supergiants, which are the main Galactic
sources of dust (Sedlmayr 1994). The location of the dust
formation window in these stars is given by the temperature
gradient in the circumstellar environment which is determined by
radiative transfer ($T \propto R^{-0.4}$, Ivezi\'{c} \& Elitzur
1997). For a typical red giant on the Asymptotic Giant Branch
this region is a shell (with a thickness of 5--10~$R_*$) lying a
few stellar radii ($R_*$) from the stellar photosphere. The
density at the inner edge of this shell is $n \sim
10^9$~cm$^{-3}$, and the pressure $P \sim 10^{-4}$~dyn
cm$^{-2}$. Shock waves induced by radial pulsations in Long
Period Variables can raise the value of the pressure by a factor
$10^2$, up to $P \sim 10^{-2}$~dyn cm$^{-2}$ (H\"ofner
1999). This greatly increases the dust grain production
(Fleischer, Gauger \& Sedlmayr 1991, 1992), and thus
the total dust-driven mass loss (from $10^{-7}$ up to
$10^{-4}$~M$_\odot$ yr$^{-1}$).  The time spent by the newly
formed grains in the dust-growing region depends on the stellar
wind velocity (10-20~km s$^{-1}$) and is typically a few years.

Figure~\ref{paths} shows the dust formation window in
pressure-temperature space, for the chemical species that lead to
the formation of dust in an O-rich (top panel) or a C-rich
(bottom panel) cool star circumstellar envelope. The thin solid
lines mark the phase transition region below which the most
important dust precursor molecules can be formed (adapted from
Lodders \& Fegley 1999).  The hatched area is the dust forming
region in the circumstellar envelope of a cool giant star. The
region is limited on the right by the thermodynamical path of a
static outflowing wind typical for an Asymptotic Giant Branch
star. The left side is obtained by increasing the maximum
pressure in the envelope by a factor 100, as during the
propagation of pulsational shocks in the atmospheres of Long
Period Variables. Dust formation in the envelopes of evolved
giants occurs in the region between the two tracks, below the
phase transition lines for each chemical species. The formation
any any individual species depends on there being sufficient time
available.

\section{Dust formation and BEL clouds}\label{belcdust}

The parameters described in the previous section define the dust
formation window in a setting where dust condensation is known to
occur copiously. We want to compare these conditions with the
ones encountered by the BEL clouds as they expand in the quasar
outflow, to evaluate their own chances of condensing dust.  The
initial conditions of the BEL clouds are very different from the
ones in cool star atmosphere. The temperature is far higher
($T_{0(BELC)} \sim 10^4$~K, Osterbrock 1989), and such a high
temperatures are hostile to dust formation.  However the density
of BEL clouds is also high ($P_{0(BELC)} \sim 0.1$~dyn cm$^{-2}$,
Osterbrock 1989), compared to the dust condensation region of
cool star atmospheres.  So if the BEL cloud expands and cools as
part of an outflowing wind, this cooling could lead them through
the dust formation window.

We will simplify the situation by assuming that the BEL cloud
(subscript {\footnotesize BELC}) material is an ideal gas obeying
polytropic adiabatic scaling, and further that the gas is
monoatomic with index $\gamma = 5/3$. This assumption is
initially good and will hold at least until the gas undergoes a
phase change of H into H$_2$, which will not occur before dust
begins to form. With this assumption we can estimate the
expansion factor required for cloud cooling below the dust
condensation equilibrium temperature:

\begin{equation}
\left( \frac{r}{r_0} \right)_{(BELC)} = \left(
\frac{T_{cond}}{T_{0\ (BELC)}} \right)^{\frac{1}{3(1-\gamma)}} 
\end{equation}

Since $(T_{cond}/T_0)_{(BELC)} \sim 0.1$, $(r/r_0)_{(BELC)} \sim
3$, which tells us that BEL clouds reach a dust formation
temperature once they have expanded by a factor of three.

Unless and until the clouds start to interact with one another,
we can approximate their pressure with the polytropic $P \propto
r^{-3\gamma}$, which gives $P_{(BELC)} \sim
10^{-3}$--$10^{-4}$~dyn cm$^{-2}$ for clouds expanded by a factor
3. A more accurate analysis of the cloud expansion process,
taking into account cloud interactions and the hydrodynamics of
the surrounding medium, will find a larger value of the
pressure. Furthermore, if we relax the hypothesis of adiabatic
expansion, allowing thermal equilibrium between the clouds and
the ionizing radiation, then the cloud temperature will drop even
faster, as their ionization parameter will rapidly decrease with
the cloud expansion (see discussion below).

The above estimates show that the BEL clouds do reach the same
values of temperature and pressure encountered in the dust
formation window of cool giant stars (Lodders \& Fegley 1999), as
shown in Figure~\ref{paths} (the thick solid lines show the
limiting case of adiabatic expansion; the lines become more
vertical as expansion is reduced). The actual formation of dust
and its precursors, and the amount of grains which are condensed,
however, depend on the time spent by the clouds in the region
favorable to grain condensation and growth. This is given by the
cloud expansion time-scale $\tau_{(BELC)} = r_0/c_0 \sim 3$~yr,
where $c_0\sim 10$~km s$^{-1}$ is the initial sound speed of the
cloud and $r_0 \sim 10^{14}$~cm (Peterson 1997) their initial
average size. Since their sound speed is similar to the wind
speed of RGB stars, this timescale is comparable with the time
spent by circum{\em stellar} grains in the region where dust growth is
most active, suggesting that the efficiency in dust production in
the BEL region and late type giant winds may be similar.

Radiatively driven winds seem to be characteristically unstable,
producing highly structured winds in stars with `P~Cygni'
profiles, OB stars and Wolf-Rayet stars (e.g. Stahl et al., 1993,
Owocki 2001), that suggest shocks may be common.  In AGB winds
these intermittent pressure enhancements boost dust production,
starting a nonlinear ``avalanche'' effect, as observed in Miras
and other Long Period Variables (Fleischer, Winters \& Sedlmayr 1999,
and references therein). Quasars show
similar structures (Turnshek 1988), which are predicted by
hydrodynamic modeling (Proga 2001), indicating that pressure
fluctuations could greatly enhance the quasar dust creation rate.

Where is the dust made? To estimate how far the clouds are from
the quasar center when the dust condensation processes turn on,
we need to couple the cloud expansion rate with their equation of
motion. We assume that the BEL clouds co-move with a warm
confining medium. With this assumption, the clouds will be
accelerated from an initial velocity $v_0 \simeq 5 \cdot 10^8$~cm
s$^{-1}$ to an asymptotic value $v_\infty \simeq 5 \cdot 10^9$~cm
s$^{-1}$. Note that the BEL clouds are moving supersonically at
velocities $\sim$100 times larger than their free expansion
rate. The ratio $u = d/d_0$ between the clouds distance at a
given time, compared with the clouds initial distance when they
detach from the quasar accretion disk, is then large. In the
limit $\beta^2 = (v_0/v_\infty)^2 \ll 1$ and $u \gg 1)$ the
equation of motion of the expanding clouds can be solved
analytically, which gives the following relation between the
expansion factor $x = (r/r_0)_{(BELC)}$ and the cloud position
$u$:

\begin{equation}
x \sim \beta \frac{\tau_0}{\tau_{(BELC)}} \ \left[ u + \frac{1}{2} \ln
u \right]
\end{equation}

\noindent
where the logarithmic term is due to the variation of the sound
speed caused by the expansion of the cloud, $\tau_0 = d_0/v_0
\simeq 24$~days is the time required for the cloud to double the
initial distance from the quasar, and $\tau_{(BELC)} \simeq 3$~yr
is the cloud expansion timescale. For $u \gg 1$ we have:

\begin{equation}\label{motion}
\frac{x}{u} \sim \beta \frac{\tau_0}{\tau_{(BELC)}} \simeq 2
\cdot 10^{-3}
\end{equation}

\noindent
which means that for $x \sim 3$ (the expansion factor at which
dust begins to condense), $u \sim 10^3$. Given an initial
distance from the quasar center of $d_0 \simeq 10^{16}$~cm (for
moderate luminosities, Peterson 1997, Sabra \& Hamann 2001), BEL
clouds will start to be dust bearing when they are $\sim$3~pc
from the quasar center.  This result also shows that the
ionization parameter $U = n_{ph}/n_e$ of the clouds becomes very
small as the clouds reach the dust forming region. Since $n_{ph}
\sim u^{-2}$ and $n_e \sim x^{-3}$, by the time the clouds are
3~pc from the the central BH, $U$ will have decreased by a factor
$3^3/(10^3)^2 \simeq 3 \cdot 10^{-5}$. This rapid decrease of the
ionization parameter will cause an even faster drop in the cloud
temperature, assuming an initial equilibrium between the clouds
and the ionizing radiation rather than pure adiabatic
expansion. As explained before, this will favor earlier dust
condensation in the clouds, with even higher efficiency due to
the higher local gas pressure.

\section{Dust survival in the BELCs}\label{survival}

Can the dust survive the radiation field from the quasar BH? The
quasar luminosity is of order 10$^{9}$ times higher than the
$10^4$~L$_\odot$ luminosity of a typical giant star. The large
flux of energetic photons from the quasar continuum may then in
principle overheat the newborn dust grains above their
sublimation temperature.  This could delay the occurence of dust
condensation until it becomes impossible, due to the ever
decreasing gas density.  However, due to the much larger
geometrical dilution in quasars, the radiative flux reaching the
BEL clouds interior is actually lower than the stellar flux in
the dust forming region of the giant's wind. For a quasar of
luminosity $10^{46}$~erg s$^{-1}$ the flux density 3~pc from the
quasar center is $\sim 10^7$~erg~cm$^{-2}$~s$^{-1}$.  This is at
least one order of magnitude less than the $\sim 2 \cdot
10^8$~erg~cm$^{-2}$~s$^{-1}$ in the stellar case. For this reason
the dust formation window of the BEL clouds is determined by the
polytropic expansion of the clouds gas, as we have assumed, and
not by radiative transfer, as in the case of circumstellar
envelopes (Ivezi\'{c} \& Elitzur 1997).

The actual flux penetrating the interior of the BEL clouds is in
fact even lower, since the neutral hydrogen and the dust
precursor molecules expected to exist in the interior of the BEL
clouds are very efficient in absorbing the quasar radiation.  In
the radial flow of Elvis (2000) the BEL clouds will stack up
radially to give a mean covering factor of unity, given their
initial 0.1 value, and so provide additional self-shielding.  The
interior of BEL clouds a few parsec from the quasar center seems
to be a safe place for dust formation, and we can conclude that
dust sublimation is prevented in the BEL clouds even for the most
luminous quasars.

Other destruction mechanisms, such as dust sputtering by
electrons and ions, or chemical sputtering, are not very
effective at the rather low ($T_K \ls 10^4$~K) kinetic
temperature of the cloud medium.  Kinetic sputtering by the
surrounding medium becomes effective (Draine \& Salpeter 1979)
only for $T_K \gs 5 \cdot 10^5$~K. By the time the BEL clouds
start forming dust, they will be surrounded by a warm medium
having a temperature $T \sim T_0 u^{2(1-\gamma)} \sim 2 \cdot
10^5$~K, which is already low enough to prevent the immediate
destruction of the nascent dust grains by sputtering.

\section{Discussion and conclusions}\label{discussion}

The composition of dust depends on the element abundances in the
originating material. If the number ratio of C/O is above unity,
then almost all the oxygen is tied up in CO which does not react
to form solids, leading to carbon rich dust; conversely if C/O is
less than unity, carbon grains are almost nonexistent, and oxides
(e.g. SiO$_2$) dominate (Whittet 1992).  Carbon is overabundant
in quasars at high redshifts (Hamann \& Ferland 1999), so
carbonaceous dust may be dominant. Quasar oxygen abundances
however, have not been well measured (Hamann \& Ferland 1999).
Carbonaceous dust formation in stars is similar to that of smoke
or soot from combustion (Frenklach \& Feigelson 1989), hence our
term `smoking quasars'.

Because dust condensation is highly nonlinear (Frenklach \&
Feigelson 1989), the amount of dust formed in any particular
quasar is hard to predict, and will depend on subtle variations
of wind conditions, including the initial density. Heavily
obscured quasars (Sanders et al., 1988) may be easily formed this
way. A comparison with unobscured quasars may pinpoint optimal
conditions for dust condensation.

If dust creation is a consequence of the quasar activity itself 
then some issues of early cosmological epochs are affected:

\begin{itemize}

\item[(1)] Quasar winds ($v>$1000~km~s$^{-1}$) readily exceed the
escape velocity from a galaxy (v$_{esc}\sim$200~km~s$^{-1}$), or
even a rich cluster of galaxies (v$_{esc}\sim$1000~km~s$^{-1}$).
Hence, unless there is a retarding force, the dust they produce
will be ejected into the intergalactic medium. Dust in the
intergalactic medium may: affect measurements of cosmological
parameters, including acceleration (Aguirre 1999); alter the
gaseous IGM composition via selective depletion of elements onto
dust grains; contribute to the far infrared background, changing
galaxy evolution models (Dwek 1986, Dwek Rephaeli \& Mather 1990,
Aguirre \& Haiman 2000); possibly alter the Sunyaev-Zel'dovich
effect (Sunyaev \& Zel'dovich 1972) in clusters of galaxies
(Marengo 2000). The potential for such effects must be enhanced
with quasars as an additional source for dust.

\item[(2)] At low redshift the quasar will effectively be
re-cycling dust from the host galaxy interstellar medium out into
the IGM, while changing the dust size distribution and
dust-to-gas ratio.  At high redshifts instead, the matter
entering the quasar accretion disk is likely to contain very
little dust, since creating large amounts of dust
(10$^{8}$-10$^{9}$M$_{\odot}$, Omont et al. 2001) at z$>$5 is
difficult to arrange (Edmunds \& Eales 1998), in part because
there is only 1~Gyr available to form red giant stars.  Quasars
can provide an additional path for dust formation, adding to the
original stock. Assuming the dust-to-gas mass ratio measured in
Long Period Variables (Knapp 1985), and noting that a quasar of
luminosity 10$^{46}$erg~s$^{-1}$ can eject up to
1~$M_{\odot}$~yr$^{-1}$ (Sabra \& Hamann 2001), a dust production
rate of $\sim$0.01~$M_{\odot}$~yr$^{-1}$ follows.

The most luminous quasars, in which the highest dust masses are
found (Omont et al. 2001), have luminosities over
10$^{47}$erg~s$^{-1}$ and so may have mass loss rates
$>$10~M$_{\odot}$~yr$^{-1}$, leading to
$\sim$10$^{7}$~M$_{\odot}$ of dust over a nominal 10$^{8}$~yr
lifetime. This mass approaches the amounts detected, although the
assumed dust-to-gas ratio is merely illustrative. As a result
the infrared emission of quasars may not require the normally
assumed large associated burst of star formation (Sanders et
al. 1988).

This is an economical explanation. If the only dust at high z is
manufactured in quasars, then far less dust/Gpc$^{-3}$ is implied
by infrared/sub-mm detections of quasars than if quasars simply
illuminate pre-existing dust, which would then need to be more
widely distributed.

\end{itemize}

In summary, dust will be created from the free expansion of
quasar broad emission line clouds in an outflowing wind. The
association of dust with quasars is not then necessarily linked
with intense star formation around quasars, but is a consequence
of the quasar activity itself. The inevitable creation of dust in
quasar winds may solve the puzzle of where the first dust comes
from, and in doing so suggests a new importance for quasars in
cosmology.


\acknowledgements 

We thank Eric Feigelson for alerting us to the problem of dust
formation at high redshift, and both him and Fabrizio Nicastro
for valuable discussions. We thank the anonymous referee for
helpful and clarifying comments.  We dedicate this paper to the
memory of Mike Penston, who taught the value of looking for the
common astrophysics of stars and quasars.  This work was
supported in part by NASA contract NAS8-39073 (Chandra X-ray
Center).


\clearpage


\figcaption[f1.eps]{Phase transition lines for O-rich (top panel)
and C-rich (bottom panel) dust precursor molecules (adapted from
Lodders \& Fegley 1999). The hatched area is the dust formation
region in the circumstellar envelopes of evolved cool giant
stars, delimited by the two cases of static and dynamic
(pulsating) AGB atmospheres. The thick solid line is the path of
BEL clouds as they expand, for two different values of their
initial density.\label{paths}}

\clearpage


\epsscale{0.8}
\plotone{f1a.eps}
\clearpage
\plotone{f1b.eps}
\clearpage

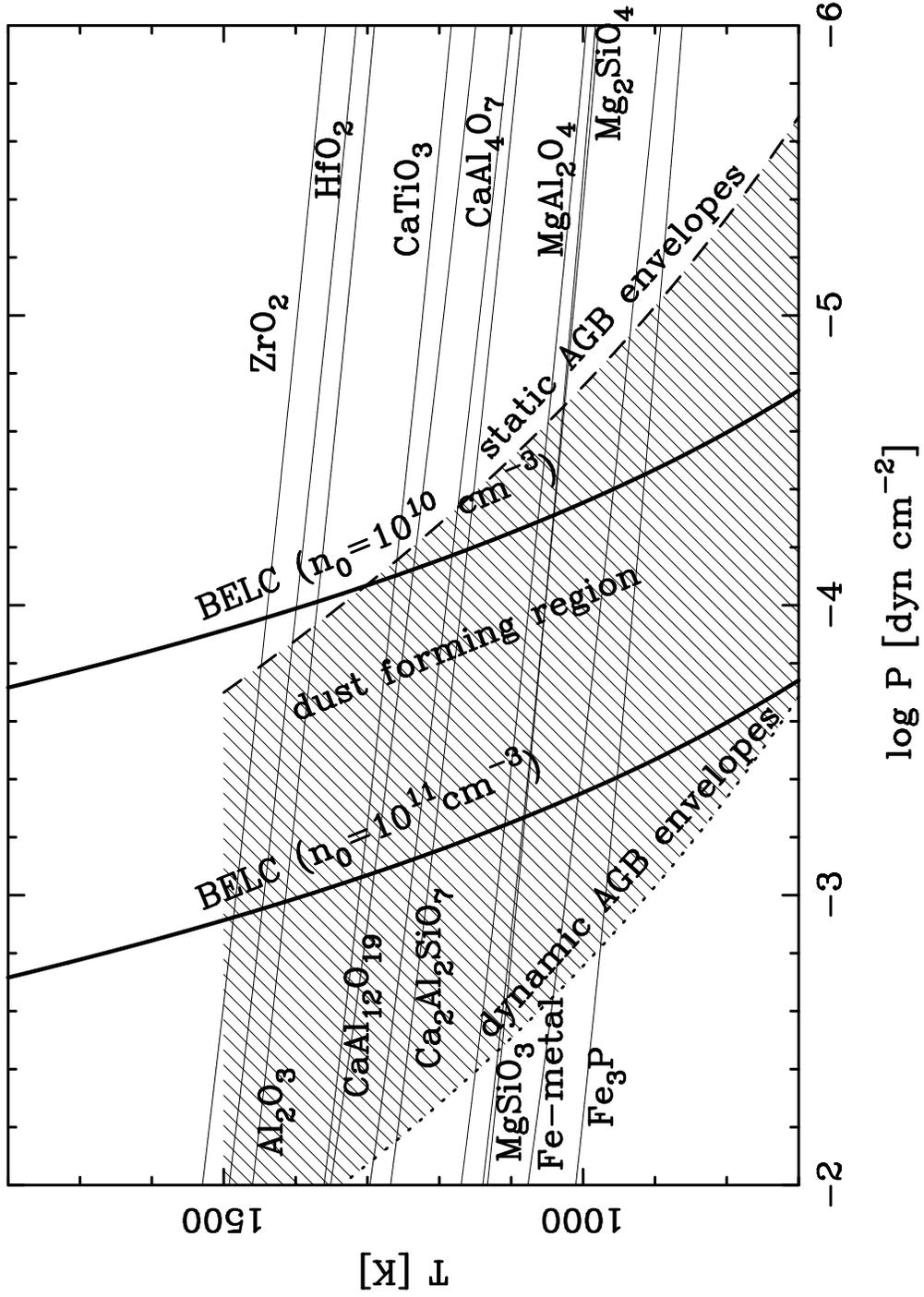
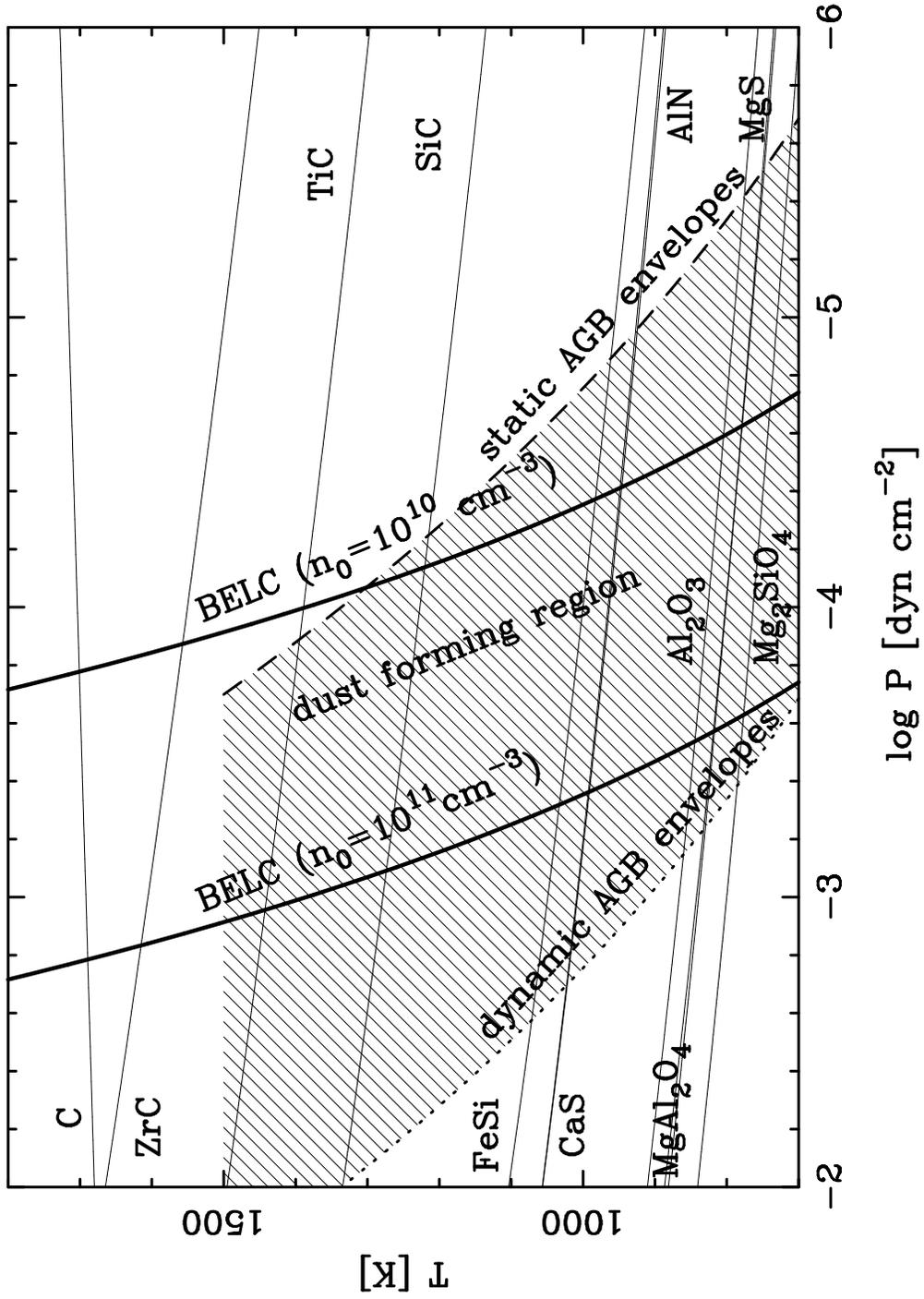
\end{document}